\documentclass[openacc]{rstransa}%%%%where rstrans is the template name

\titlehead{Perspective}

\begin{document}

\title{Evolution of solar and stellar coronal abundances due to magnetic activity}

\author{
David H. Brooks$^{1,2,3}$, Deborah Baker$^{2}$, \\
David M. Long$^{3}$, Paola Testa$^{4}$ \\
and Harry P. Warren$^{5}$}

\address{$^{1}$Computational Physics, Inc., Springfield, VA 22151, USA\\
$^{2}$University College London, Mullard Space Science Laboratory, Holmbury St. Mary, Dorking, Surrey, RH5 6NT, UK\\
$^{3}$National Institutes of Natural Sciences, National Astronomical Observatory of Japan, 2-21-1 Osawa, Mitaka, Tokyo, 181-8588, Japan\\
$^{4}$Centre for Astrophysics \& Relativity, School of Physical Sciences, Dublin City University, Glasnevin, D09 K2WA, Dublin, Ireland\\
$^{5}$Harvard-Smithsonian Center for Astrophysics, 60 Garden St, Cambridge, MA 02193, USA\\
$^{6}$Space Science Division, Naval Research Laboratory, Washington, DC 20375, USA}

\subject{astrophysics, stars, spectroscopy}


\corres{David H. Brooks\\
\email{dhbrooks.work@gmail.com}}

\begin{abstract}
We discuss the evolution of solar coronal element abundances over an active region lifetime.
Magneto-convection drives the complexity of magnetic fields 
that emerge above the photosphere. This complexity is dissipated, together with that of the overlying pre-existing fields, through dynamic events such as flares.
A period of stable "ordinary" coronal heating ensues, before the concentrated fields are dissipated through interactions with the surrounding environment.
The evolution of coronal abundances can be explained by the First Ionisation Potential (FIP) effect operating within this framework.
We extend the discussion from magnetic activity on timescales of active region lifetimes (months),
to the solar cycle (years), and stellar evolution (eons). The broad picture shows intriguing similarities that may prompt new investigations.
\end{abstract}

%\maketitle

\begin{fmtext}
\end{fmtext}
\maketitle

\section{Introduction}
The First Ionization Potential (FIP) effect is now established as an increase in the abundance of low FIP elements (such as Fe, Mg, Si) in the
solar corona compared to the photosphere, while the high FIP elements (such as Ne and O) retain their photospheric abundances or are slightly depleted. 
The boundary between low- and high-FIP elements is usually observed to be around 10\,eV. 
The effect was originally noted from spectroscopic observations of the solar
corona during
rocket flights in the 1960s \cite{Pottasch1963}, and was later detected in the solar wind, solar energetic particles (SEP), and galactic cosmic ray sources \cite{Meyer1985}.
The effect is thought to operate in the chromosphere 
where there are many potential ionizing processes. H I Ly$\alpha$ radiation is dominant, however, and the transition energy of 10.2\,eV is close
to the observed boundary between low- and high-FIP elements. It therefore ionizes the low FIP elements only and creates a separated reservoir of ions 
and neutrals. A mechanism is then needed to preferentially transport the ions to the corona, that is likely intimately
linked to the coronal heating process itself, and leads to an observed fractionation of elements in the corona. 

A promising possibility is the ponderomotive force (PF) arising from propagating Alfv\'{e}n waves. These may be generated in the convection zone \cite{McIntosh2011a} or produced by magnetic
reconnection \cite{Laming2004,Laming2015}. Alfv\'{e}n waves propagating downward from the corona reflect and refract after impinging on the chromosphere from above. The
associated upward ponderomotive force transports ions to the corona.
Different fractionation scenarios that modulate the magnitude of the FIP effect and/or the behaviour of different element ratios result from whether the waves are resonant on closed field structures or non-resonant on open field 
\cite{Laming2019,Mihailescu2023}. This is potentially useful for understanding compositional differences between the fast solar wind,
that is thought to originate in open field regions such as coronal holes where non-resonant waves are pervasive, and the slow solar wind that may originate in coronal loops. Recent observational
evidence supports both the idea of a role for wave activity in FIP fractionation \cite{Stangalini2021,Baker2021} and the existence of wave reflection in the chromosphere \cite{Murabito2024}. The chromosphere is, of course, a complex and dynamic environment. Recent multi-fluid magnetohydrodynamic (MHD) simulations including collisional coupling and the effects
of wave damping suggest that ponderomotive acceleration could occur in the direction of wave propagation \cite{Martinez2023}. Further work is needed to fully understand 
the FIP effect in realistic dynamic solar conditions.

Another aspect of the FIP effect that has generated recent interest is the detection of the inverse-FIP (iFIP) effect on the Sun \cite{Doschek2015} during flares. An inverse effect
was first observed on RS CVn-type binary stars \cite{Brinkman2001,Drake2001} 
and some very active K-dwarfs \cite{Gudel2001}. Subsequent observations found that M-dwarfs also show an iFIP effect \cite{Robrade2005}. 
Some doubts about the initial observations were founded on the uncertainties associated with the 
photospheric abundances of solar-like stars, but further independent support has come from the detection of the iFIP effect in localized regions of strong magnetic
field, such as sunspots or sunspot light bridges, during solar flares \cite{Doschek2015,Baker2019}. The iFIP effect has also now been detected in the slow solar 
wind \cite{Brooks2022}. 

\begin{figure}[!h]
\centering\includegraphics[width=\textwidth]{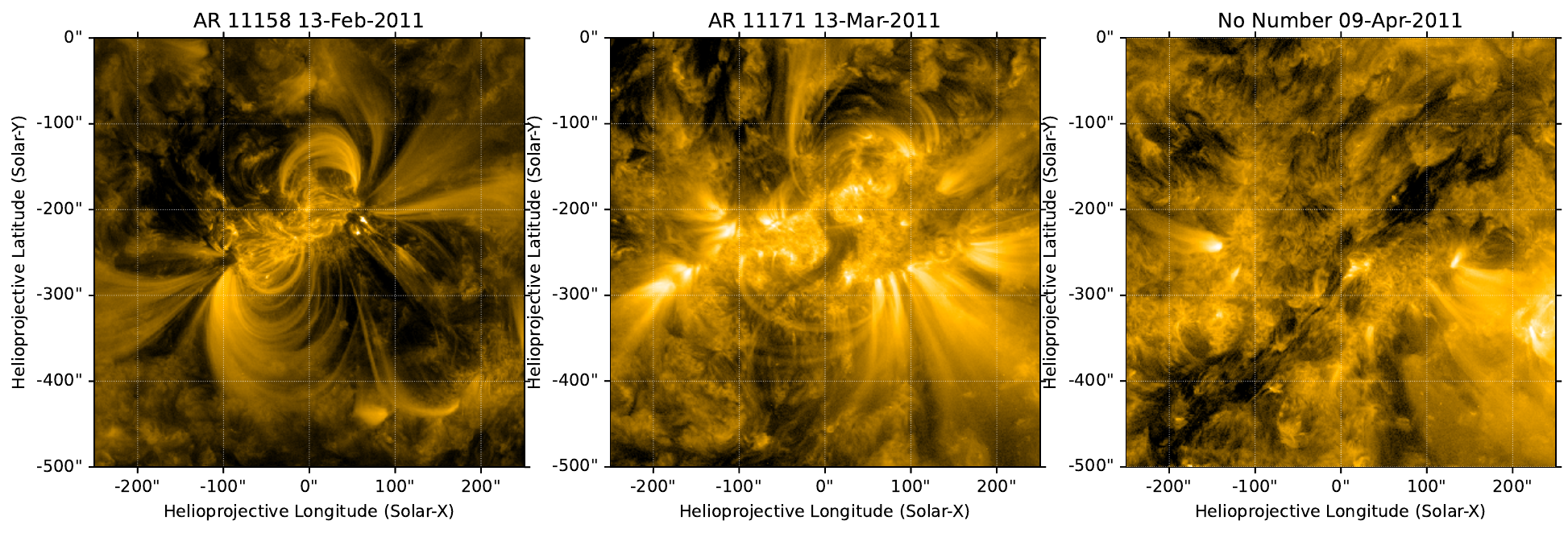}
\caption{SDO/AIA 171\,\AA\, images of the evolution of AR 11158/11171 as it passes the central meridian on three rotations. The magnetically complex, dynamic, and compact
structure of emerging AR 11158 develops into the larger, diffuse, non-flaring AR 11171, and finally devolves into a filament channel before dissipation in the quiet Sun.}
\label{fig1}
\end{figure}

A key question is whether the iFIP effect is a unique clue to the workings of the coronal heating mechanism, or whether the FIP and iFIP effects are 
just different aspects of the same basic process. The models of the FIP effect we have referred to here \cite{Laming2004,Martinez2023} both invoke the
ponderomotive force acting preferentially on the low FIP ions with the iFIP effect arising from the direction of the ponderomotive force, which is
related to the direction of Alfv\'{e}n wave propagation in these models. For example, if
fast mode waves generated by subchromospheric reconnection are propagating up from below, the ponderomotive force will be directed downwards in one of the models \cite{Laming2004}.
There is now some observational
evidence that both effects work on the low FIP elements. The detection of the iFIP effect in the slow solar wind using particle counting techniques
indicated a depletion of the low FIP element
Fe \cite{Brooks2022}, while the extended cooling times of iFIP patches in post-flare coronal loops suggests reduced radiative losses also indicative of low FIP element
depletion \cite{Brooks2018}. The same conclusion was earlier drawn from measurements of column depths in iFIP locations \cite{Doschek2016}. 
Of course, the observations are complex and contradictory, and in theory the iFIP effect could be a distinct process enhancing neutrals. There is also 
evidence that the FIP effect is caused by the depletion of high FIP elements in some stellar coronae \cite{Peretz2015}.

These developments highlight the complexity of the 
FIP and iFIP effects arising potentially from the direction of wave propagation (upward/downward), and resulting in disparate behaviour (enhancement/depletion)
in different elements (low and high FIP). Nonetheless, the recent observations and theoretical work suggest that
\textit{the overarching process is simply the depletion of low FIP elements from the chromosphere} with the different observed (FIP/iFIP) effects resulting from the opposite direction of
diffusion. Future theoretical work will pin down whether the depletion is 
\textit{in the direction of the
wave source} (e.g. coronal or sub-chromospheric reconnection), or \textit{in the direction of wave propagation} (e.g. aligned with spicules).

In light of this background, we review the known results on the evolution of coronal abundances in active regions (ARs), and develop a picture of the changes due
to magnetic activity. 
In this paper, when we refer to AR coronal loops, we are referring to what we observe in EUV and/or soft X-ray images of the solar corona: 
a narrow emission feature where we can observe the footpoints and the apex. 
There is some evidence that observed loops are composed of much finer threads that are below the resolving power of current instrumentation, 
but this is a question that has not been settled \cite{Klimchuk2006,Warren2008,Brooks2012}.
Following the description of AR evolution, we extend
the discussion to solar cycle timescales, and the solar-stellar connection. 

\section{Coronal abundance evolution in active regions}
While our description of the evolution of coronal abundances in ARs is intended to be general, we refer to one widely studied region
(AR 11158) to aid the discussion.
To observe all the features in such an AR comprehensively throughout its lifetime is challenging with the limited field-of-view (FOV) spectrometers that can measure plasma composition.
Nevertheless, here we assume that the activity in this AR is broadly illustrative of a typical AR lifetime and can be related to the spectroscopic properties measured in other ARs. 

AR 11158 emerged around 2011 February 10, and is shown Figure \ref{fig1} after substantial growth on February 13.
This region produced multiple flares and coronal mass ejections and has been the subject of intensive studies \cite{Tarr2013,Inoue2013,Kay2017,Norton2021}.
The AR grows rapidly from $\beta\rightarrow\beta\gamma\rightarrow\beta\gamma\delta$ Hale class producing an M6.6 flare a couple of days after emergence, and
reaches a total area in excess of 3.04$\times$10$^{19}$\,cm$^2$ and total unsigned magnetic flux of at least 4.22$\times$10$^{21}$\,Mx as early as February 12 \cite{Warren2012}.
The region was observed as it tracked across the disk, and then again on the second rotation as AR 11171, and the subsequent rotation as an unnumbered diffusing
active area. Figure \ref{fig1} shows images of the AR as it crosses the central meridian on the three rotations. 
AR 11158 emerged into a relatively quiet environment - although there is some 
interaction with AR 11156 to the solar west - and initially consisted of two bipoles before developing into a highly complex region with four sunspots
and strong shearing of the neutral line. Such magnetic complexity, driven by magnetoconvection below the surface, always leads to significant flare activity as the magnetic stresses are relaxed.
AR 11158 produced one X-class, 7 M-class, and 66 C-class flares according to the Hinode Flare Catalog \cite{Watanabe2012}. Activity reduced to a single M1.5 flare as AR 11171 on the second rotation, and no flares on the third rotation.

Hinode/EUV Imaging Spectrometer (EIS) \cite{Culhane2007} data for this region show that the emission measure distribution is peaked near 3.5\,MK in its early emergence phase, 
but that the slopes above and below the
peak are shallower than found in magnetically stronger ARs, suggesting relatively lower frequency heating in the core \cite{Warren2012}. This is presumably because it was observed by EIS while
still in the emergence phase, when it was clearing the surrounding environment and its own magnetic complexity. Lower frequency heating in the core at this time
is in agreement with observations of high variability at the footpoints of hot core loops in other ARs \cite{Testa2014,Testa2020,Cho2023}.
Observations have also shown that the peak temperature decreases and enhanced cool and hot emission below and above the temperature peak dissipates with time as ARs age i.e. the heating frequency 
in the AR core increases \cite{Ugarte2012}. 

If a magnetically simple and stable AR emerges into a quiescent environment, the AR corona will be heated through processes that
might be described as "ordinary" coronal heating. Of course the actual heating mechanism is unknown, but if we adopt the definition, discussed earlier, of a coronal loop as a distinct feature
in an EUV image that is composed of a number of sub-resolution magnetic threads \cite{Warren2008,Brooks2012}, then examples of these processes might include 
internal reconnection of braided threads within the loops, such as occurs in nanoflares \cite{Parker1988} or in an MHD avalanche \cite{Hood2016,Cozzo2023},
or alternatively, high frequency Alfv\'{e}n wave turbulence heating \cite{VanB2011}, or chromospheric jets supplying mass and energy
and forming loop structures \cite{DePontieu2009,DePontieu2011,Martinez2017,Bose2025}. 
The idea is that a specific {\it internal} heating process will occur within the AR loops, and this will naturally occur if the AR evolution is undisturbed by the
pre-existing surrounding magnetic environment. 

Conversely, if the AR and pre-existing magnetic environment is highly complex, then the AR evolution will first be affected by {\it external} 
reconnection with pre-existing fields, dissipation of emerging twisted 
magnetic fields, and/or large scale field misalignments \cite{Reale2019,TestaReale2023}.
This dissipation of magnetic complexity in the early phase by flares and flare-like intermediate processes
\cite{Reale2019,TestaReale2023} ultimately dominates the initial AR evolution, and delays the AR from 
settling down to the ordinary coronal heating phase seen in magnetically simpler ARs. 
Eventually the AR region field weakens and starts to interact with surrounding quiet Sun fields. This might 
lead to similar signatures as observed in the early phase, albeit of less dramatic magnitudes, for example, small-scale impulsive heating events rather than larger scale flaring events. 
Magnetically complex ARs can thus pass through three broad phases:
\begin{itemize}
\item Emergence - dominated by external processes that remove complexity  (reconnection with pre-existing field, dissipation of twist)
\item Stable - dominated by internal processes (ordinary coronal heating)
\item Dissipation - competition between internal processes (ordinary coronal heating) and external interactions (with the surrounding magnetic environment).
\end{itemize}
With this picture in
mind, 
the elemental abundances that we measure should reflect a transition from a complex field external reconnection based development
to an internal heating mechanism controlled development. 
Here we review what plasma composition measurements we might expect to make, and what we actually observe. 

During the early emergence phase one would expect to see photospheric or weakly fractionated abundances indicative of emerging flux \cite{Baker2018}, essentially
reflecting the composition of the quiet corona \cite{Lanzafame2005}. When flares are
dominant, 
we see a wide variety of results depending on flare-class, flare-phase, and elements measured. Large flares lead to the evaporation of
photospheric material, at least at 10\,MK \cite{Warren2014}, though measurements show spatial variations from photospheric to coronal 
abundances in features such as post-flare loops \cite{Doschek2018}. Coronal material has also been found in the flare decay phase in X-ray spectra \cite{Mondal2021,Rao2023}.
A recent paper gives a 
summary of many of the flare composition results in its Appendix \cite{To2024}.
Close to the strong magnetic fields of sunspots we may also see
the iFIP effect. We also see a large amount
of impulsive heating events on a smaller scale in the transition region. They are predominantly clustered around the neutral line where large flares are also expected
\cite{Brooks2008}. These events also show a photospheric composition \cite{Warren2016}.
The key point is that at this time a whole range of large and small-scale flaring composition signatures are competing to dilute the characteristic
abundances of the AR in its stable phase.
When the external reconnection has stopped, the AR quickly forms a very typical structure, with a 
hot core (3--4\,MK) above the moss at the footpoints of high temperature loops, peripheral high lying warm (2\,MK) loops, and bright fan loops (1\,MK) at the AR edges intermixed 
with high temperature outflows that have been connected with the solar wind \cite{Sakao2007,Harra2008,Doschek2008}. 
On the subject of the outflows, they also form quickly and again
there are arguments that this is an external reconnection induced development. Observations of one AR showed that within hours the typical AR structure was established with the reconfiguration of the magnetic field that formed the outflows apparently
initiated by a small CME (coronal mass ejection) \cite{Brooks2021}. 
Interchange reconnection \cite{Fisk2003,Crooker2012} between closed and open field at the AR boundary is a popular explanation for the 
formation of the outflows. In EIS observations, the predominant abundance diagnostic uses Si and S lines \cite{Feldman2009,Brooks2011}, and the outflows show a strong
FIP effect \cite{Brooks2011,Testa2023}.

The closed magnetic field structures in the AR core also show a strong FIP effect in measurements using the same Si/S diagnostic \cite{Warren2012}. The underlying moss would be expected to reflect a similar composition, though
there is also some evidence of mixing due to dynamic events \cite{Baker2013,Brooks2020,Testa2023}. Warm loops are assumed to show coronal abundances, but there are relatively few
good measurements of these loops. In EIS observations, the presumed coronal
abundance leads to relatively weak S emission, which makes measurements difficult. The bright fan loops certainly show a strong FIP effect \cite{Young1997,Warren2016}. 

This then is the basic compositional structure of ARs. 
As they age, however, further changes occur. Studies from Skylab initially indicated that the FIP bias (ratio of coronal to photospheric abundances) increases \cite{Widing2001}
but this does not seem to be observed clearly with EIS. EIS observations have shown that the regions dissipate through interactions with the surrounding environment, including eruptive activity \cite{Baker2015,Baker2018}.

Considering this perspective on this typical AR evolution, one could ask what we would see in an averaged AR spectrum rather than in spatially resolved features. 
The Skylab results found in the early phase of the ARs evolution, when we expect flares to occur, could be understood in that we 
would measure emerging flux with a photospheric composition, and this flux reconnects with pre-existing field to cause flaring, some fraction of which also produces
photospheric composition. This photospheric composition might obscure the appearance of the coronal composition signature associated with the AR emerging in the quiet Sun until all the complexity has gone and the AR has moved to a stable phase driven by ordinary coronal heating.
In this scenario
it is not so much that the FIP bias is increasing, but that we initially observe the dominance of structures and events that have a photospheric FIP bias, and later we then observe 
the structures that have a coronal composition dominating. This dominance of coronal composition plasma then increases
as the AR grows and develops with more and more of the AR encompassing the surrounding area. Finally, there would be a reversed phase at the end of the AR lifetime
as the region dissipates, although observations of such dissipating ARs are rare. To be clear, the emergence$\rightarrow$stable$\rightarrow$dissipation cycle we discuss above would
manifest in the composition measurements as dominated by photospheric (emerging flux/flares)$\rightarrow$coronal (AR)$\rightarrow$mixed (dissipated AR/quiet Sun).

\begin{figure}[!h]
\centering\includegraphics[width=0.5\textwidth]{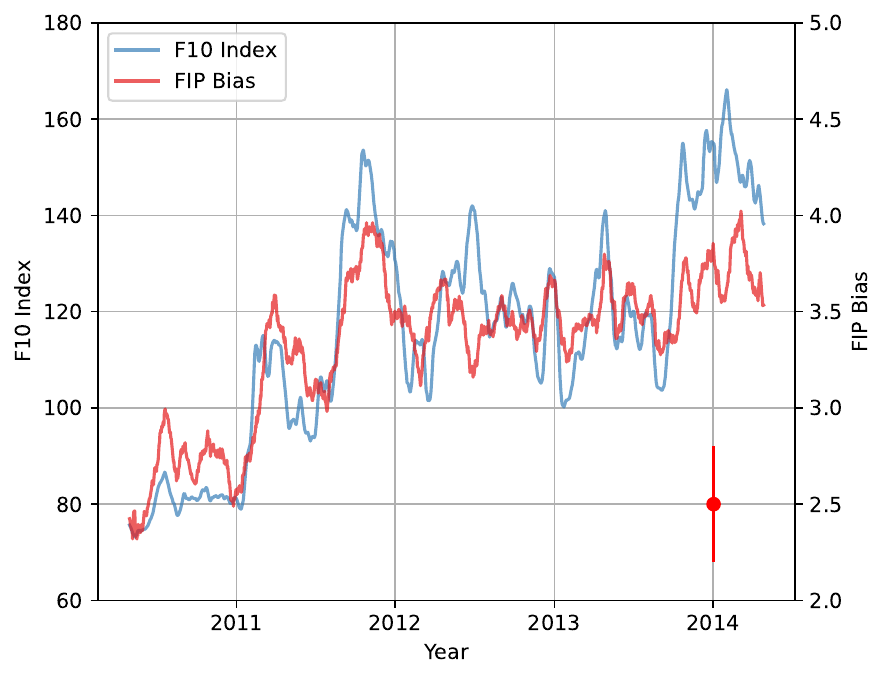}
\caption{Temporal evolution of the full disk integrated F10.7\,cm radio flux (blue) and the FIP bias (red) between April 2010 and May 2014. The data are 27-day Carrington rotation running averages. The FIP bias is measured by computing the full disk temperature distribution (DEM) using spectral lines of the low FIP elements (Mg, Si, Fe) observed by 
SDO/EVE and modelling the intensity of a strong line from the high FIP element Ne. The ratio of predicted to observed intensity yields the FIP bias. 
This plot is an adaption of Figure 2 from Brooks et al. (2017) \cite{Brooks2017} and their error estimate is plotted as the red vertical bar. }
\label{fig2}
\end{figure}

\section{Coronal abundance evolution over the solar cycle}
If our picture of elemental abundance changes associated with AR evolution is close to being correct, it naturally leads us to consider what we would see on solar cycle timescales.
Elemental abundances in the solar wind are known to show variations with the solar cycle \cite{Kasper2007,McIntosh2011b,Lepri2013,Kasper2012,Shearer2014}.
The high FIP element He, for example, increases in abundance at solar maximum in the slow wind \cite{McIntosh2011b}.
An important question is whether this reflects cyclic variations in the source regions, or only reveals itself in the solar wind after acceleration. 
Evidence from coronal streamer and Sun-as-a-star observations suggests the former explanation \cite{Landi2015,Brooks2017}. Figure \ref{fig2} shows
the variation of FIP bias and F10.7\,cm (2.8\,GHz) radio flux during the rise phase of solar cycle 24 from April 2010 to May 2014. The FIP bias measurements are from
the Extreme ultraviolet Variability Experiment (EVE) on SDO and reflect the variation of the coronal abundances of the low FIP elements Fe, Mg, and Si, compared to the high FIP element Ne. The FIP bias is
strongly correlated with the F10.7\,cm radio flux during the rising phase ($r$ = 0.9), though it appears to saturate at solar maximum. 
One uncertainty is that the result is based on the high FIP element Ne, and the O/Ne ratio shows a similar increase at solar maximum
in the solar wind \cite{Shearer2014}. This behaviour could reflect the lower FIP of O relative to Ne, or could be due to a depletion of Ne. 
The issue is discussed further in the original paper \cite{Brooks2017}.

This solar cyclic dependence of full disk integrated coronal abundances is broadly explained by the contribution of ARs (and other features) to the averaged spectrum.
At solar minimum, there are few if any ARs, and a typical quiet Sun coronal composition is measured. In Figure \ref{fig2} it is 2.3, though the F10.7\,cm radio flux has already risen 
above the absolute solar minimum value, so if the correlation holds throughout the solar cycle the FIP bias may go lower. This will also depend on the abundance 
diagnostic used. EIS Si/S and Fe/S measurements suggest a value closer to 1.4 or 1.5 \cite{Ko2016}, though this is influenced by the behaviour of S, which fractionates differently
in different circumstances \cite{Laming2019}. In fact EIS spatially resolved observations of the full Sun allow us to disentangle the contributions of quiet features and 
ARs to the measured FIP bias of 1.4. In specific observations from January 2013 \cite{Brooks2015}, 
the area coverage of ARs with FIP bias in the range 2--4 is $\sim$28\%. The mean value when combined
with the $\sim$72\% coverage of features with an average FIP bias $\sim$1 is close to 1.4. In that context the EVE results look a little high. This could reflect the different
elements used, but the calculation is also different.
The FIP bias comes from full disk average spectra, so it is the contribution of the spectral line intensities rather than the area 
coverage that matters. Consider, for example, the EIS observations of the strong Fe XII 195.119\AA\ line. A typical value in the quiet Sun is $\sim$135 erg cm$^{-2}$ s$^{-1}$ 
\cite{Brooks2009}, whereas it was measured as $\sim$2729 erg cm$^{-2}$ s$^{-1}$ in AR 11158 \cite{Warren2012}. For an EVE observation with ARs like AR 11158 covering 
30\% of the area of the disk the ARs contribute 90\% of the total intensity of the spectrum. 

During the rise phase of the cycle, the area coverage of ARs
increases and their contribution to the full disk integrated measurement consequently increases. In this phase, the ARs are isolated and magnetically simple. They emerge
with a photospheric or weakly fractionated composition, but this signature would be hard to detect in the full Sun data even if there were several ARs on disk. 
Since they are magnetically simple and are emerging into a less magnetically complex environment, 
they proceed to the ordinary (internally driven) coronal heating phase relatively quickly and develop a coronal composition. This signature is seen in the rise of the
FIP bias with the cycle, and it increases as ARs of larger magnetic flux appear and as the area coverage increases. 

As we approach solar maximum, more complex
ARs appear, but their composition signatures now include the initial phase where their complexity is dissipated by external processes inducing large flares and 
impulsive events that evaporate photospheric,
or even iFIP effect, plasma. They develop a strong FIP effect as they transition to the ordinary coronal heating phase, but the overall competition between the 
signatures of many ARs at different stages of their evolution leads in part to the saturation of FIP bias observed in Figure \ref{fig2}. 
This also manifests itself as a non-linear component in the FIP bias-F10.7\,cm radio flux correlation at high levels of solar activity, 
which can be seen as the separation of the curves in Figure \ref{fig2} around 2014.

Another reason we observe 
the non-linear component during
increased magnetic activity is that there is a tendency for gyroresonance emission to increase its contribution to the radio flux when there are
larger, stronger magnetic concentrations, whereas the FIP bias tends to be lower above sunspots than it is in more evolved ARs \cite{To2023}. 
It may be that a clearer linear correlation could be observed by comparing the FIP bias with the total unsigned magnetic flux with the sunspot contribution removed, but
this has not been attempted yet.
Large equatorial coronal holes that form at solar maximum also complicate the picture. EIS observations indicate that at least some equatorial coronal holes show
a FIP effect \cite{Brooks2015}.

\section{Coronal abundance evolution of the Sun in time}
There have been extensive studies of solar-like stars that have attempted to relate FIP bias measurements in stellar coronae to properties of the stars. 
Some of these investigations have revealed a dependence on spectral type, suggesting a relationship with fixed properties such as rotation rate and/or surface temperature
\cite{Wood2010}. It is clear that there is also some dependence on activity level as measured by, for example, the X-ray to bolometric luminosity ratio \cite{Testa2015}. 

\begin{figure}[!h]
\centering\includegraphics[width=0.5\textwidth]{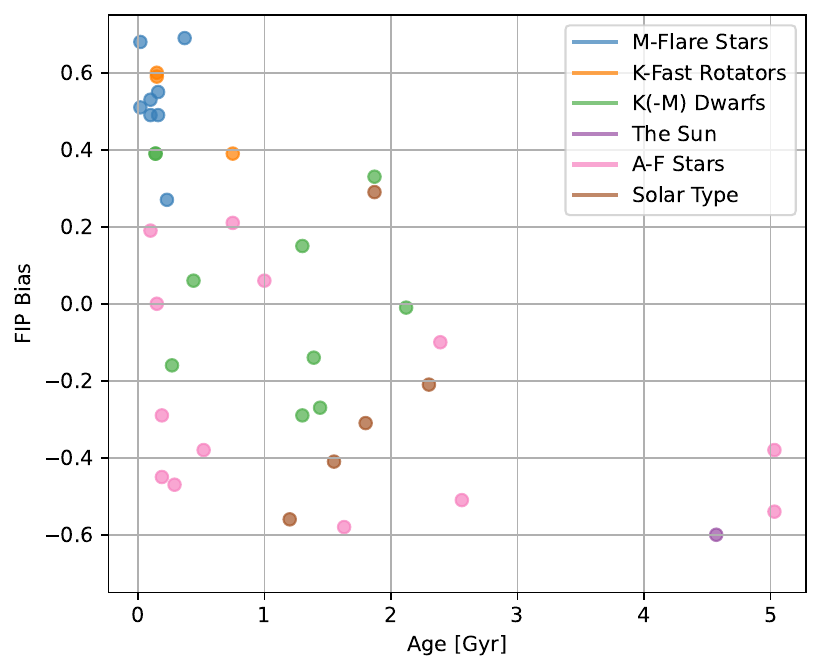}
\caption{FIP bias against stellar age for a sample of magnetically active stars \cite{Seli2022}. We plot a subset of the original literature sample
excluding evolved RS CVn-type stars, those with undetermined ages, and Proxima Centauri. 
The definitions of the stellar type in the legend are taken from Table 1 in the original paper. The ages and FIP bias values
are taken from Table 2. We use the "literature" FIP bias values, which are derived from measurements of the coronal to photospheric ratio
of the mean abundances of some combination of C, N, O, and Ne relative to Fe i.e. [X/Fe] = $\log$[X/Fe]$_{cor}$-$\log$[X/Fe]$_{phot}$.
This definition means values of [-0.5,0,0.5] correspond to a solar FIP bias of [0.32,1,3.2]. The active Sun appears at 4.57\,Gyr. 
Stellar measurements of photospheric abundances were used when available, but
otherwise solar photospheric abundances from \cite{Asplund2009} and \cite{Drake2005} for Ne were used. 
The specific values are given in Table A.1 of the Appendix of the paper \cite{Seli2022}. 
}
\label{fig3}
\end{figure}

If we relate our discussion of the solar cyclic FIP bias effect (based on our discussion of AR composition evolution) to the development
of solar-like stars on stellar evolutionary time-scales, we might expect a relationship between FIP bias magnitude and stellar age, and this has been
investigated in the past \cite{Telleschi2005}. This would be the 
case if the coronae of magnetically complex and active stars were analogous to the Sun's corona at the peak of the solar cycle phase, and if the coronae of more evolved and less active stars
were analogous to the Sun's corona at solar minimum. 
Of course, such comparisons are difficult to make due to the number of variables involved: different stellar classes are not necessarily comparable,
stellar masses, rotation rates, metallicities etc. are different. In fact, it is not the case:
young highly active stars show a strong iFIP effect that dominates their coronae. We can see examples
of this in
Figure \ref{fig3}, which shows the FIP bias as a function of age for a sample of magnetically active stars \cite{Seli2022}. 
This study explored the dependence of the FIP bias on many of the stellar properties (surface gravity, metallicity, age, Rossby number etc.)
while prioritising homogeneity in the parameters in the sample. They also attempted to address the significant uncertainty in similar studies
that results from adopting solar photospheric abundances in the FIP bias calculation by using (or in some cases measuring) 
the stellar photospheric abundance. From their results reproduced in Figure \ref{fig3}, we can see that
flaring M-dwarf stars, and some K-dwarfs, especially fast rotators,
show a strong iFIP effect. Most of these stars have estimated ages of $<$ 2\,Gyr. In contrast, A-F and solar type stars show a normal coronal FIP bias or close to photospheric abundances. 
Stars older than 3\,Gyr show a solar-like FIP effect. 

The evolutionary change could be from an iFIP effect to a solar-like FIP effect \cite{Telleschi2005}. This
can be understood within our discussion framework. We know from solar observations that the iFIP effect occurs in magnetically complex ARs and so far has only been seen
in high temperature emission during solar flares. It has also been associated
with e.g. subchromospheric reconnection \cite{Baker2024}. Stars that show an iFIP effect, therefore, may have coronae with emission dominated by highly complex ARs that can produce 
iFIP effect plasma. External
reconnection processes such as flare reconnection, or subsurface interactions of sunspots, are known to occur in such complex ARs. 
We should recognize, however, that once this complexity is cleared on evolutionary timescales (i.e. the magnetic activity of the star has decreased), 
the corona should come to be dominated by 
ARs in their ordinary, stable (internally driven) coronal heating phase, or indeed somewhat more magnetically simple ARs such as are seen in the early rise phase of
the solar cycle. Basically, stars with an iFIP effect corona should 
move directly from the top left to the lower half of Figure \ref{fig3} without passing through a phase of photospheric dominance, regardless of their later evolution.
Stars with coronal emission dominated by photospheric abundances 
could just be at the minimum phase of their cycles, or all activity may have ceased, or they may never have developed a FIP effect in their coronae if their chromospheres are relatively
inactive \cite{Drake1995,Wood2010}.

\section{Discussion}
We have attempted to draw together connections between observations of coronal abundance evolution over a solar ARs lifetime, the solar cycle, and stellar evolution timescales.
We put forward an overview of the influence of magnetic activity on coronal composition on these timescales. It is clear, however, that there are 
fundamental complexities that we have necessarily simplified. 

At a basic level, all of the elemental abundance measurements are founded on EUV and X-ray spectroscopic techniques that are built on a number of assumptions. The differential emission measure (DEM)
analysis that underpins the measurements has long been a source of controversy and there is a view that the techniques are fundamentally limited \cite{Craig1976,Judge1997,Testa2012}. It is easy to generate
conflicting abundance results using different DEM methods even when the underlying observational and atomic data are basically the same. Furthermore,
in these evolving magnetic environments the impact of plasma dynamic effects such as non-equilibrium ionization, non-Maxwellian electron distributions, 
collisional distruption of dielectronic recombination and ionisation, can all play a role. An integrated study of the competition and influence of these effects on abundance
measurements still awaits. 

Accepting the results as they are, we see systematic differences dependent on the species used in the analysis. We mentioned the behaviour of elements such as S that lie close to the boundary
between low and high-FIP classifications, but models and observations also show different levels of fractionation between different low-FIP or high-FIP element pairs \cite{Laming2019}. We see also
differences in FIP bias that are dependent on spatial location, such as the differences from footpoint to loop top in post-flare loops \cite{Doschek2018,To2024} and sigmoidal ARs \cite{Baker2013}.

On the stellar evolution arguments, as noted some of the measurements are predicated on the assumption of solar photospheric abundances since stellar photospheric data
are not always available, and the evolution of coronal
abundances over the solar cycle also implies that any of the stellar data points in Figure \ref{fig3} could be moving.
Nevertheless, we believe that the overarching picture we have described can help to highlight areas that need further investigation in the future.

\vskip6pt
\ack{We thank the Royal Society for its support throughout the organization of the Theo Murphy meeting and the production of this special issue. The work of DHB and HPW was supported by the NASA Hinode program. The work of PT was supported by NASA contract NNM07AB07C (Hinode/XRT) to the Smithsonian Astrophysical Observatory, contract 8100002705 (IRIS) to the Smithsonian Astrophysical Observatory, NASA Heliophysics Guest Investigator grant 80NSSC21K0737, and by the NASA Heliophysics Supporting Research grant 80NSSC21K1684.
The AIA and EVE data are courtesy of NASA/SDO and the AIA, EVE, and HMI science teams.
}

\end{document}